\documentclass[12pt]{article}

\usepackage[utf8]{inputenc}
\usepackage[american]{babel}

\usepackage{palatino}
\fontfamily{ppl}\selectfont
\usepackage{setspace}
\usepackage[letterpaper, margin=1in]{geometry}

\usepackage{amsmath}
\usepackage{amssymb}
\usepackage{dcolumn}

\usepackage{graphicx}
\usepackage{float}
\usepackage{subcaption}
\usepackage{tikz}
\usetikzlibrary{calc,matrix,positioning,shapes.geometric,decorations.pathreplacing}
\usepackage{pdflscape}
\usepackage{lscape}
\usepackage{placeins}

\usepackage{booktabs}
\usepackage{multirow}
\usepackage{multicol}
\usepackage{threeparttable}
\usepackage{longtable}
\usepackage{tabularx}
\usepackage{adjustbox}

\usepackage{natbib}
\usepackage{hyperref}
\hypersetup{
    colorlinks=true,
    linkcolor=blue,
    citecolor=blue,
    urlcolor=blue
}

\usepackage[colorinlistoftodos]{todonotes}
\usepackage{verbatim}
\usepackage{import}
\usepackage{titling}

\newcommand{\hypmark}[1]{\hypertarget{#1}{}}

\title{Help Converts Newcomers, Not Veterans: Generalized Reciprocity and Platform Engagement on Stack Overflow}

\author{
  Lenard Strahringer\thanks{Corresponding author: \href{mailto:lenardst@stanford.edu}{lenardst@stanford.edu}} \\
  \small Stanford University
  \and
  Sven Eric Pruess \\
  \small University of Sydney \& University of Münster
  \and
  Kai Riemer \\
  \small University of Sydney
}

\date{\today}

\begin{document}

{\singlespacing
\maketitle
\thispagestyle{empty}

\begin{abstract}
\noindent Generalized reciprocity — the tendency to help others after receiving help oneself — is widely theorized as a mechanism sustaining cooperation on online knowledge-sharing platforms. Yet robust empirical evidence from field settings remains surprisingly scarce. Prior studies relying on survey self-reports struggle to distinguish reciprocity from other prosocial motives, while observational designs confound reciprocity with baseline user activity, producing upward-biased estimates. We address these empirical challenges by developing a matched difference-in-differences survival analysis that leverages the temporal structure of help-seeking and help-giving on Stack Overflow. Using Cox proportional hazards models on over 21 million questions, we find that receiving an answer significantly increases a user's propensity to help others, but this effect is concentrated among newcomers and declines with platform experience. This pattern suggests that reciprocity functions primarily as a contributor-recruitment mechanism, operating before platform-specific incentives such as reputation and status displace the general moral impulse to reciprocate. Response time moderates the effect, but non-linearly: reciprocity peaks for answers arriving within a re-engagement window of roughly thirty to sixty minutes. These findings contribute to the theory of generalized reciprocity and have implications for platform design.

\end{abstract}
}

\newpage

\section{Introduction}

Whether receiving help on an online platform motivates subsequent helping behavior has been a persistent question in the study of online communities. Generalized reciprocity, the tendency to help others after being helped oneself \citep{nowak_upstream_2007, baker_paying_2014}, is among the mechanisms proposed to sustain cooperation on platforms like Stack Overflow and Wikipedia, where direct exchange between the same two individuals is rare. If receiving help reliably increases the probability of giving it, cooperation may sustain itself through cascading helping acts, even among strangers \citep{fowler_cooperative_2010}. The behavioral evidence, however, is inconsistent: some observational studies find substantial reciprocity effects \citep{baker_paying_2014, chen_why_2019}, while others find weak or null relationships \citep{wasko_why_2005, joyce_predicting_2006}. We argue that much of this inconsistency is methodological.

Observational studies of reciprocity on platforms face a persistent confound: more engaged users are more likely both to seek help and to provide it, regardless of whether their own questions are resolved. Studies that correlate help received with help given, without accounting for this baseline activity, conflate reciprocity with general user engagement \citep{baker_paying_2014, chen_engaging_2018, joyce_predicting_2006}. A related limitation is that prior designs have largely treated receiving help as a one-time event, even though help-seeking and helping occur continuously.

We address both problems by developing a matched difference-in-differences survival analysis, applied to over 21 million questions from Stack Overflow. For each question a user posts, we define observation windows before and after the question and compare changes in helping behavior between users whose question received an answer (treated) and propensity-score-matched users whose question did not (control). We use Cox proportional hazards models to account for the continuous-time dynamics of helping behavior and the staggered, time-varying structure of treatment.

We develop two theoretical predictions alongside the basic reciprocity hypothesis. Drawing on theories of socialization and norm internalization, we predict that the reciprocity effect is stronger among newcomers than among experienced users (H2). Newcomers have not yet internalized community-specific norms or learned to value platform incentives such as reputation points; general moral obligations to reciprocate are therefore more likely to drive their behavior. As users accumulate experience, these general-purpose triggers may be displaced by community-specific motivations. We also predict that faster responses should amplify the reciprocity effect (H3), because gratitude is more intense when help arrives while the need is still salient \citep{mccullough_is_2001, tsang_gratitude_2006}.

The results support H1, H2, and broadly H3. Receiving an answer increases the hazard of helping by roughly 6\% in the pooled sample (HR\,=\,1.06, 95\%~CI\,=\,[1.04, 1.07]). This effect declines from HR\,$\approx$\,1.09 among users with less than one week of tenure to a null effect among those with more than six years. Speed interactions are consistently negative across experience groups: faster responses produce stronger reciprocity, consistent with H3. One wrinkle: in the pooled discrete-bin analysis, the effect is near zero for responses arriving under 30 minutes, peaks at 30--60 minutes (HR\,$\approx$\,1.18), and attenuates at longer delays, reflecting a limited window where users can be re-engaged by receiving help that we describe in the results.

Our study makes three contributions. First, we provide evidence that receiving help on Stack Overflow increases subsequent helping, using an identification strategy that addresses the endogeneity problems present in prior work. Second, we show that this effect is concentrated among newcomers and substantially attenuated among experienced users, consistent with the view that reciprocity functions primarily as a contributor-recruitment mechanism rather than a driver of sustained helping. Third, we identify a re-engagement window in which platform session dynamics moderate the reciprocity effect, qualifying the direct application of laboratory-based theories of emotional immediacy to task-driven field settings.

\section{Theory}

\subsection{Generalized Reciprocity}

A central mechanism proposed to support cooperation in online communities is \textit{generalized reciprocity}, the tendency for individuals who have received help to help others in turn, even when the recipient is not the original benefactor \citep{nowak_upstream_2007, fowler_cooperative_2010}. Unlike direct reciprocity, which involves returning favors to the same individual, generalized reciprocity requires no repeated interactions or exchange with the same person. Instead, it operates as a pay-it-forward model, sustaining collective contributions within groups where members may have infrequent or non-recurring interactions, such as online knowledge-sharing platforms.

Socio-psychological research identifies two complementary mechanisms underlying generalized reciprocity: moral reasoning and emotional responses. On the moral side, individuals may be guided by the logic of universalization \citep{levine_logic_2020}: they consider what would happen if all members requested more help than they gave. Such deliberation typically leads individuals to conclude that widespread free-riding would undermine the community as a valuable source of help, generating a shared moral obligation to reciprocate \citep{curry_is_2019}. This renders the relationship between individual and group into an indirect exchange: if one receives help from the group, one feels obligated to help the group in return.

Emotional mechanisms reinforce these moral considerations. A prominent response is gratitude, aroused by the perceived generosity of others \citep{mccullough_is_2001, mccullough_adaptation_2008}. Gratitude has been shown to motivate prosocial actions toward others beyond direct reciprocation \citep{tsang_gratitude_2006, bartlett_gratitude_2006}. According to \citet{molm_building_2007}, such emotional responses are especially relevant for systems where free riding is easy, such as online knowledge-sharing platforms. Experiencing gratitude in these contexts not only increases the propensity to help but also strengthens perceptions of the community as valuable and worth sustaining \citep{elster_norms_2009}.

Generalized reciprocity is among the most widely studied behavioral mechanisms explaining cooperative behavior in groups, owing to its potential to initiate cascading acts of generosity \citep{fowler_cooperative_2010, boyd_evolution_1989}. Comparative research and meta-analyses indicate that it is a globally shared moral principle \citep{curry_is_2019}. Robust experimental evidence comes from both laboratory \citep{greiner_indirect_2005, molm_building_2007, stanca_measuring_2009} and field settings \citep{mujcic_indirect_2018, fowler_cooperative_2010}.

\subsection{Empirical Challenges in Measuring Generalized Reciprocity}

Despite its theoretical importance, empirical findings on generalized reciprocity in online communities are mixed, and we argue that this inconsistency is largely attributable to persistent methodological challenges. Understanding these challenges is essential both for interpreting existing findings and for motivating our analytical approach.

A first challenge concerns the reliance on self-reported motives in surveys and interviews. Contributors frequently describe reciprocity as a central reason for their engagement \citep{wasko_it_2000, lakhani_how_2003}, yet sociological and psychological research has demonstrated that people have difficulty articulating the moral motives behind their actions \citep{vaisey_motivation_2009, mcclelland_how_1989}. When users state that they help because they enjoy it or feel a sense of obligation, it is unclear whether this reflects genuine reciprocity, altruism, conformity to perceived social norms, or the popular rhetoric of 'giving back' as a post-hoc rationalization. The distinction between reciprocity and other prosocial motives, such as the intrinsic joy of helping or the pursuit of social status, is particularly difficult for individuals to make through introspection \citep{batson1991altruism, benabou_incentives_2006}. As a result, self-reported data might overstate the prevalence of reciprocity as a discrete motive.

A second and more fundamental challenge is the endogeneity of help-seeking and help-giving in observational data. Users who ask questions on platforms like Stack Overflow are also disproportionately likely to answer questions, not because receiving help triggered a reciprocal response, but because both behaviors reflect the same underlying characteristics, such as greater platform engagement. Studies that correlate help received with help given without accounting for this baseline activity effectively confound reciprocity with user type \citep{baker_paying_2014, chen_engaging_2018, joyce_predicting_2006}. As we discuss below, the confound between reciprocity and user type is the central identification challenge that motivates our matched diff-in-diff design.

\subsection{Reciprocity and the User Trajectory}

We propose that the mixed findings in the literature can be partly reconciled by attending to a dimension that prior research has largely overlooked: the user trajectory from newcomer to experienced contributor. We argue that generalized reciprocity is strongest for new users and attenuates as users gain experience on the platform.

The logic of this prediction draws on theories of socialization and norm internalization. When a user first encounters an online community, they are unfamiliar with its specific norms, social structures, and reward systems. At this stage, community-specific incentives, such as reputation points, badges, or leaderboard positions, carry little motivational weight because the newcomer has not yet learned to value them. Instead, initial contributions are guided by more general moral principles that operate independently of community-specific knowledge. Generalized reciprocity is precisely such a principle: it requires only that the individual has received help and feels a moral obligation or emotional impulse to help in return, without any need to understand or identify with the particular community \citep{levine_logic_2020, curry_is_2019}.

As users interact with the community, they gradually learn its norms and develop a sense of membership. On platforms like Stack Overflow, these norms are codified in explicit reputation systems that reward helpful behavior with visible status markers. Once users internalize these community-specific norms, the broad moral obligation of reciprocity may be displaced by more localized incentive structures. The logic of universalization (``I should help because I was helped'') gives way to the logic of community membership (``I should help because that is what valued members do''). This norm displacement has been theorized in related contexts by \citet{benabou_incentives_2006}, who show how external incentives can crowd out moral motives.

Several additional mechanisms accelerate this decline. First, as users accumulate experiences of receiving help, each additional instance becomes less socially surprising and less likely to elicit strong gratitude \citep{mccullough_adaptation_2008}. The emotional response that fuels reciprocity habituates over time. Second, as users observe the platform's incentive structure that rewards helpful behavior with reputation and status, they may increasingly attribute others' helpful behavior to status pursuit rather than genuine generosity, further dampening the gratitude that triggers reciprocal helping.

This account predicts that generalized reciprocity is strongest among newcomers and declines with experience. This also helps explain why prior studies have reached different conclusions about the importance of reciprocity: those that focus on or oversample experienced users may underestimate reciprocity, while those that capture newcomer behavior may find stronger effects.

\hypmark{hyp:1}
\begin{quote}
\textsc{Hypothesis 1 (H1).} \emph{Receiving help increases a user's propensity to help others (generalized reciprocity effect).}
\end{quote}

\hypmark{hyp:2}
\begin{quote}
\textsc{Hypothesis 2 (H2).} \emph{The generalized reciprocity effect is strongest among newcomers and decreases as users gain experience on the platform.}
\end{quote}

\subsection{Response Time and the Immediacy of Reciprocity}

The emotional mechanisms underlying reciprocity suggest that the speed of help matters for the strength of reciprocity. Gratitude, the primary emotional driver of generalized reciprocity \citep{mccullough_is_2001}, is an affective response that is triggered by the perception of a benefactor's generosity. The intensity of this emotional response is likely to depend on the characteristics of the helping experience, including its timeliness.

A swift response to a help request can amplify the experience of receiving help through several channels. First, a fast answer signals that the community is attentive and generous, strengthening the perception that the group is worth sustaining through reciprocal contributions. Second, gratitude is an emotional response that may be most intense when it is temporally proximate to the perceived need. When a user posts a question, their need for help is salient; a quick answer arrives during this window of heightened need. The longer the user waits for help, the less valuable the help is and the more likely they are to have gotten help elsewhere. Thus, quick responses are likely to intensify feelings of gratitude.

Conversely, delayed responses may attenuate reciprocity. When an answer arrives hours or days after the question was posted, the user's original need may have faded; they may have found a solution elsewhere, or the problem may have become less pressing. The interval between asking and receiving weakens the associative link between the two events, potentially diluting the emotional response. Furthermore, longer waits may signal that the community is less responsive or generous, weakening the perception that underpins the moral obligation to reciprocate.

These considerations lead to a straightforward prediction:

\hypmark{hyp:3}
\begin{quote}
\textsc{Hypothesis 3 (H3).} \emph{Faster responses to a user's question strengthen the generalized reciprocity effect, while slower responses weaken it.}
\end{quote}

We note that this hypothesis follows from the dominant theoretical accounts of generalized reciprocity, which emphasize emotional immediacy and the salience of the helping experience. However, as we will show, field settings introduce structural dynamics, particularly the behavioral consequences of users being online and offline, that qualify the predictions of laboratory-based theory. We return to this issue when presenting our results.

\section{Empirical Setting: Stack Overflow}
\label{sec:empirical_setting}

Stack Overflow offers a valuable empirical setting for examining generalized reciprocity in online knowledge sharing. The platform provides publicly available quarterly data dumps containing rich longitudinal information on user behavior, including knowledge-seeking and knowledge-providing activities dating back to the platform's inception. This extensive dataset enables the reconstruction of helping patterns over time, allowing for the analysis of temporal sequences of receiving and providing help, which is critical for studying generalized reciprocity. We note that Stack Overflow activity has declined in recent years, partly due to the rise of generative AI tools; however, the core question-and-answer dynamics we study are shared broadly across help forums and knowledge-sharing communities, supporting the generalizability of the mechanisms we identify.

At its core, Stack Overflow is a question-and-answer platform for software developers and technologists. As of mid-2025, the site had accumulated over 24 million questions and 36 million answers from more than 4.5 million users. Its user base spans a wide range of experience levels and professional roles in software development and learners \citep{stack_overflow_stack_2024}. Questions on the platform must be tightly focused on specific programming problems; broad or opinion-based inquiries are typically closed by community moderation \citep{stack_exchange_inc_how_2024}. The most frequently discussed topics, as reflected in question tags, include JavaScript, Python, Java, C\#, PHP, Android, and HTML, spanning areas from web development and mobile applications to data science and machine learning. Stack Overflow is built around several core features: tags categorize questions by topic, facilitating expertise matching; a voting system allows users to upvote or downvote questions and answers to assess quality; and reputation points are earned for valued contributions such as providing high-quality answers; \citep{stack_exchange_inc_how_2024}.

For our study, the key feature is the question-and-answer interaction between users. When a user posts a question, other users may provide answers. This creates a clearly timestamped sequence (question posted, answer received) that we exploit to measure the temporal dynamics of generalized reciprocity. Crucially, on Stack Overflow, the identities of question-askers and answerers are typically different individuals, making this an ideal context for studying \textit{generalized} reciprocity: users who receive help from one community member may subsequently help entirely different members.

\section{Data and Methods}
\label{sec:methods}

To test our predictions, we develop a matched staggered difference-in-differences (DiD) survival analysis that leverages the temporal structure of help-seeking and help-giving on Stack Overflow. Our design compares changes in helping behavior (measured as the hazard of posting an answer) before and after an user receives an answer, between users who receive an answer and matched users who do not. This approach represents a novel integration of two methodological traditions: the staggered difference-in-differences framework of \citet{callaway_santanna_2021}, which accommodates variation in treatment timing and treatment effect heterogeneity, and the counting-process formulation of the Cox proportional hazards model \citep{andersen_gill_1982}, which allows treatment status and other covariates to vary continuously over the observation window.

\subsection{Study Design}
\label{sec:study_design}

For each question posted by a user, we construct a timeline centered on two key events: the posting of the question ($T_{\text{question}}$) and, for questions that receive an answer, the time at which the first answer with a non-negative vote score is posted ($T_{\text{answer}}$). We define an observation window spanning the two days before the question is posted ($T_{\text{question}} - 2\text{D}$) through the two days after ($T_{\text{question}} + 2\text{D}$). This yields a 4-day timeline for each question, divided into distinct phases.

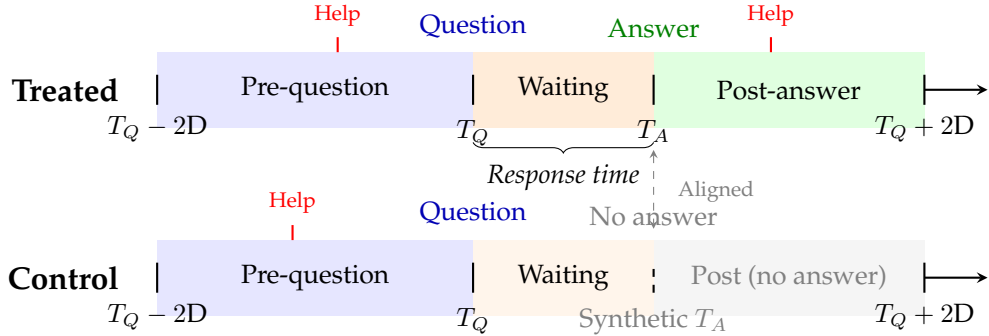
\begin{figure}
\caption{Study Design: Matched Difference-in-Differences Survival Analysis}
\label{fig:study_design}
\centering
\begin{tikzpicture}[x=1.2cm, y=1cm, >=stealth, font=\small]
\node[anchor=east, font=\bfseries] at (-0.3, 2) {Treated};
\draw[thick, ->] (0,2) -- (9.2,2);
\fill[blue!10] (0,1.5) rectangle (3.5,2.5);
\fill[orange!15] (3.5,1.5) rectangle (5.5,2.5);
\fill[green!12] (5.5,1.5) rectangle (8.5,2.5);
\draw[thick] (0,1.8) -- (0,2.2);
\draw[thick] (3.5,1.8) -- (3.5,2.2);
\draw[thick] (5.5,1.8) -- (5.5,2.2);
\draw[thick] (8.5,1.8) -- (8.5,2.2);
\node[below, font=\footnotesize] at (0,1.8) {$T_Q - 2\text{D}$};
\node[below, font=\footnotesize] at (3.5,1.75) {$T_Q$};
\node[below, font=\footnotesize] at (5.5,1.75) {$T_A$};
\node[below, font=\footnotesize] at (8.5,1.8) {$T_Q + 2\text{D}$};
\node[font=\footnotesize] at (1.75, 2) {Pre-question};
\node[font=\footnotesize] at (4.5, 2) {Waiting};
\node[font=\footnotesize] at (7.0, 2) {Post-answer};
\draw[decorate, decoration={brace, amplitude=4pt, mirror}] (3.5, 1.3) -- (5.5, 1.3);
\node[below, font=\footnotesize\itshape] at (4.5, 1.15) {Response time};
\node[above, font=\footnotesize, text=blue!70!black] at (3.5, 2.55) {Question};
\node[above, font=\footnotesize, text=green!50!black] at (5.5, 2.55) {Answer};
\node[anchor=east, font=\bfseries] at (-0.3, -0.5) {Control};
\draw[thick, ->] (0,-0.5) -- (9.2,-0.5);
\fill[blue!10] (0,-1) rectangle (3.5,0);
\fill[orange!8] (3.5,-1) rectangle (5.5,0);
\fill[gray!8] (5.5,-1) rectangle (8.5,0);
\draw[thick] (0,-0.7) -- (0,-0.3);
\draw[thick] (3.5,-0.7) -- (3.5,-0.3);
\draw[thick, dashed] (5.5,-0.7) -- (5.5,-0.3);
\draw[thick] (8.5,-0.7) -- (8.5,-0.3);
\node[below, font=\footnotesize] at (0,-0.7) {$T_Q - 2\text{D}$};
\node[below, font=\footnotesize] at (3.5,-0.75) {$T_Q$};
\node[below, font=\footnotesize, text=gray] at (5.5,-0.75) {Synthetic $T_A$};
\node[below, font=\footnotesize] at (8.5,-0.7) {$T_Q + 2\text{D}$};
\node[font=\footnotesize] at (1.75, -0.5) {Pre-question};
\node[font=\footnotesize] at (4.5, -0.5) {Waiting};
\node[font=\footnotesize, text=gray] at (7.0, -0.5) {Post (no answer)};
\node[above, font=\footnotesize, text=blue!70!black] at (3.5, 0.05) {Question};
\node[above, font=\footnotesize, text=gray] at (5.5, 0.05) {No answer};
\draw[dashed, gray, <->] (5.5, 1.2) -- (5.5, 0.15);
\node[right, font=\scriptsize, text=gray] at (5.65, 0.65) {Aligned};
\draw[red, thick] (2.0, 2.5) -- (2.0, 2.7) node[above, font=\scriptsize, text=red] {Help};
\draw[red, thick] (6.8, 2.5) -- (6.8, 2.7) node[above, font=\scriptsize, text=red] {Help};
\draw[red, thick] (1.5, 0) -- (1.5, 0.2) node[above, font=\scriptsize, text=red] {Help};
\end{tikzpicture}

\vspace{0.5em}
\parbox{\textwidth}{\footnotesize\textit{Note.} Each observation window spans 4 days centered on the focal question ($T_Q$). The treated timeline (top) shows three phases: the pre-question phase (baseline), the waiting period (question posted, no answer yet), and the post-answer phase (after the first qualifying answer $T_A$ arrives). The control timeline (bottom) is from a propensity-score-matched user whose question did not receive an answer. To ensure temporal alignment, the control's post-question period is split at a synthetic transition point matching the treated user's response time. Red markers indicate help events (answers posted by the focal user to other users' questions). The treatment effect is the differential change in helping hazard between treated and control observations in the post-answer phase, relative to the pre-question baseline. The staggered timing of $T_A$ across observations motivates our use of matched DiD.}
\end{figure}

We define three phases within each observation window (see Figure~\ref{fig:study_design}). The first phase, the \textit{pre-question phase}, is from the start of the observation window to the posting of the question ($T_{\text{question}} - 2\text{D}$ to $T_{\text{question}}$). This phase captures baseline helping activity before the focal question is asked.
The second phase, the \textit{waiting period} is from the posting of the question to the receipt of the first answer ($T_{\text{question}}$ to $T_{\text{answer}}$). During this interval, the user has asked for help but has not yet received it. 
The final \textit{post-answer phase} is from the receipt of the answer to the end of the observation window ($T_{\text{answer}}$ to $T_{\text{question}} + 2\text{D}$). This is the phase in which the reciprocity effect, if it exists, should manifest.

A key feature of our design is that the post-answer (treated) phase begins at different elapsed times across observations, depending on how quickly the posted question receives a response.

This staggered timing of the post-answer phase corresponds to the setting studied by \citet{callaway_santanna_2021}, who show that standard two-way fixed-effects estimators can be biased when treatment timing varies, and treatment effects are heterogeneous. Our approach avoids this pitfall by explicitly modeling the time-varying change in treatment status within each observation window and by comparing treated observations only to not-yet-treated or never-treated controls.

For control users (those whose questions did not receive an answer), there is no natural transition between the waiting period and the post-answer phase. To maintain temporal alignment with the matched treated observation, we assign the control observation a synthetic phase transition at the same elapsed time as the treated user's response time. That is, if the treated user's answer was posted 4 hours after their question, the matched control observation's timeline is split at $T_{\text{question}} + 4\text{h}$, creating an aligned ``post'' phase of equal duration. This ensures that any differences between treated and control observations in the post-answer phase are not driven by differences in the length of the observation period.

The survival clock in our Cox models is centered on posting of the respective question: time zero corresponds to the start of the observation window ($T_{\text{question}} - 2\text{D}$), so that the question itself is posted at day~2 of the timeline and the observation window closes at day~4. All help events (answers posted by the focal user) are recorded with their exact timestamps and converted to elapsed time on this clock.

Within each timeline, we track all instances in which the user provides an answer to another user's question. These helping events constitute the outcome in our survival analysis. Conceptually, our design asks: Does the hazard of helping increase after a user receives an answer, relative to the change in hazard for a matched user who does not receive an answer? We present descriptive evidence on the raw help rates over this timeline in Section~\ref{sec:results} (Figure~\ref{fig:help_rate_pooled}), before turning to the formal Cox model estimates.

\subsection{Data}

We utilized the full sample of non-deleted questions from Stack Overflow's inception on 31 July 2008 through 1 April 2025. We addressed two edge cases that would render our fixed-length observation window invalid. First, for early platform questions, the pre-period could predate the platform's launch, meaning no answers were possible at that time. We removed 1,080 such cases. Second, due to the data dump's April~1, 2025, cut-off, the post-period could be prematurely truncated. We excluded 8,029 cases affected by this. We then excluded 2,671,121 questions that received an answer from the question poster. These filtering steps resulted in a pre-matching dataset of 21,033,429 questions from 4,764,048 unique users. After propensity score matching (described below), the final analytical dataset comprises 24,537,194 matched question-level observations (some control questions appear more than once) from 3,872,841 unique users, with a 50\% treatment rate by construction.

\subsection{Variables}

Table~\ref{tab:variables} provides an overview of the variables used in our analysis. The phase terminology is consistent with Figure~\ref{fig:study_design}: the pre-question phase captures baseline behavior, the waiting period spans the interval from question to answer, and the post-answer phase captures the period after the answer arrives. We use ``post-question'' to refer to the combined waiting period and post-answer phase whenever distinguishing between these two sub-periods is not necessary.

\begin{table}
\caption{Overview of Variables}
\label{tab:variables}
\centering
\begin{tabularx}{\textwidth}{@{}lX@{}}
\toprule
\textbf{Variable} & \textbf{Description} \\
\midrule
\multicolumn{2}{@{}l}{\textit{Outcome (Survival Event)}} \\
\hspace{1em} helpEvent & A binary event indicator equal to 1 at times when the user posts an answer to another user's question within the observation window. \\
\midrule
\multicolumn{2}{@{}l}{\textit{Time-Varying Covariates}} \\
\hspace{1em} PostQuestion & Binary; 1 for intervals occurring after the user posts their question ($t \geq T_{\text{question}}$), 0 otherwise. Captures the combined post-question period (waiting period + post-answer phase). \\
\hspace{1em} PostAnswer & Binary; 1 for intervals occurring after the user receives their first answer ($t \geq T_{\text{answer}}$), 0 otherwise. For control users (no answer received), this variable switches on at the synthetic transition point at the matched treated user's response time. Isolates the post-answer phase. \\
\hspace{1em} treatedPostQuestion & Interaction of treatment $\times$ phasePostQuestion; captures differential behavior of treated users in the entire period after the question. \\
\hspace{1em} isTreatedActive & Interaction of treatment $\times$ phasePostAnswer; the primary treatment effect capturing the additional lift in helping hazard after receiving an answer, beyond any change attributable to the act of asking a question. \\
\hspace{1em} ResponseTime & Time between question and the first answer (excluding down-voted answers) of the treated question. We include interactions with all four variables above. \\
\bottomrule
\end{tabularx}
\end{table}

\paragraph{Outcome.} The event of interest in our survival analysis is the posting of an answer. Within each observation window, we track when (and whether) the focal user provides an answer to any other user's question. The Cox model estimates the instantaneous hazard of this helping event as a function of time-varying covariates that capture the phase of the observation timeline and the treatment status.

\paragraph{Treatment.} The treatment is the receipt of an answer. We operationalize this using \textit{treatment}, a binary variable indicating whether the user's question received at least one answer with a non-negative score. We adopt this operationalization rather than Stack Overflow's ``accepted answer'' mechanism to address endogeneity: marking an answer as accepted requires the user to return to the site, which may itself correlate with helping behavior.

\paragraph{Response Time.} For each treated observation, the response time variable used in the interaction terms is $\log(1 + t_{\text{answer}})$, where $t_{\text{answer}}$ is the elapsed time in hours from the start of the observation window ($T_Q - 2\text{D}$) to the arrival of the first answer with a non-negative vote score. Because $T_Q - 2\text{D}$ is a fixed origin common to all observations, this is a monotone transformation of the actual response time in hours ($t_{\text{answer}} - t_{\text{question}}$) and preserves the full ordering of fast versus slow answers; however, it shifts the scale by the constant pre-question window length. For numerical stability, both response-time interaction terms ($\text{isTreatedActive} \times \log(1+t_{\text{answer}})$ and $\text{treatedPostQuestion} \times \log(1+t_{\text{answer}})$) are clipped to their 5th--95th percentile range and then standardized (z-scored) before model fitting. Consequently, the reported interaction coefficients $\gamma$ and $\delta$ are expressed per one standard deviation of the respective interaction term rather than per unit of $\log(1+t_{\text{answer}})$.

\paragraph{User Experience.} To test Hypothesis 2, we stratify our analysis by user tenure, defined as the number of days between a user's first activity on the platform and the start of the observation window. We define seven tenure buckets using right-closed intervals: $[0, 7]$, $(7, 30]$, $(30, 180]$, $(180, 365]$, $(365, 1095]$, $(1095, 2190]$, and $(2190, \infty)$ days, corresponding to the labels less than one week, one week to one month, one to six months, six to twelve months, one to three years, three to six years, and more than six years, respectively. Running separate models by bucket enables us to examine how the reciprocity effect varies across the user trajectory without imposing a parametric functional form. 

\subsection{Propensity Score Matching}
\label{sec:matching}

Because the receipt of an answer (the treatment) is correlated with user characteristics (more experienced users, for instance, might tend to ask higher-quality questions that are more likely to receive answers), we employ propensity score matching \citep{rosenbaum_rubin_1983} to improve the comparability of the treatment and control groups. Importantly, matching is performed \textit{across different users}: each treated question (from a user who received an answer) is matched to a control question (from a different user whose question did not receive an answer) that has a similar propensity to receive an answer based on observable pre-treatment characteristics. This cross-user matching ensures that treated and control observations are independent conditional on the matched covariates, while the DiD structure within each pair absorbs residual differences.

We estimate the propensity score using the following pre-treatment covariates, which jointly capture user experience, recent activity, and question characteristics as depicted in Table \ref{tab:covariates}. It should be noted that \textit{tagAnswerRateAvg} is critical for addressing the concern that question complexity may confound the treatment: questions in tags with low answer rates are inherently more difficult, and matching on this rate ensures that treated and control questions come from similarly tractable topic areas.
Matching on the tag-level answer rate is particularly important for our response time analysis, as it mitigates the concern that slower response times simply proxy for differently sized sub-communities with lower throughput. We use nearest-neighbor matching with a caliper of 0.05 on the propensity score.

\begin{table}
\caption{Definition of Matching Covariates}
\label{tab:covariates}
\centering\small
\begin{tabular}{p{4cm} p{9cm}}
\toprule
\textbf{Variable name} & \textbf{Definition} \\
\midrule
\textit{Exact matching} & \\
calendarYear & Calendar year exact matching indicating the year of the observation, controlling for secular trends in platform activity and answer rates. \\
topLevelTag & First tag of the question indicating the main topic and community that sees the question \\

\multicolumn{2}{l}{\textit{Propensity score matching}}\\
userTenure & Number of days since the user's first recorded activity on the platform, capturing cumulative experience. \\
numQuestionsAskedAT & Total number of questions asked by the user prior to the observation window. \\

numHelpProvidedAT & Total number of answers provided by the user prior to the observation window. \\

numQuestionsAsked30D & Number of questions asked in the 30 days preceding the observation window. \\

numHelpProvided30D & Number of answers provided in the 30 days preceding the observation window. \\

numQuestionsAsked7D & Number of questions asked in the 7 days preceding the observation window. \\

numHelpProvided7D & Number of answers provided in the 7 days preceding the observation window. \\

tagAnswerRateAvg & Average share of questions that receive an answer within the question's tags, capturing topic-specific difficulty. \\
\bottomrule
\end{tabular}
\end{table}

Table~\ref{tab:balance} displays the covariate balance before and after matching. Before matching, several covariates showed a meaningful imbalance between the treated and control groups. For instance, treated users had asked more questions in the prior 30 days (SMD\,=\,0.197) and 7 days (SMD\,=\,0.174), reflecting that users who ask questions that attract answers tend to be more recently active. The largest imbalance was in the tag-level answer rate (SMD\,=\,0.695), confirming that question topic characteristics strongly predict answer receipt and underscoring the importance of matching on this variable. After nearest-neighbor propensity score matching, all standardized mean differences fell well below the conventional threshold of 0.1, with the largest residual imbalance at 0.029 (see Table~\ref{tab:balance}). This confirms adequate balance across all matching covariates, and the balance is also illustrated in the Love plot in Appendix~\ref{sec:balance_plot}.

\begin{table}
\caption{Covariate Balance Before and After Matching}
\label{tab:balance}
\centering\small
\adjustbox{max width=\textwidth}{%
\begin{tabular}{lcccccc}\toprule
 & \multicolumn{3}{c}{Unmatched} & \multicolumn{3}{c}{Matched} \\ \cmidrule(lr){2-4} \cmidrule(lr){5-7}
 Covariate & Tr Mean & Ct Mean & SMD & Tr Mean & Ct Mean & SMD \\\midrule
 userTenure & 570.95 & 720.62 & $-$0.167 & 569.58 & 568.20 & 0.002 \\
 numQuestionsAskedAT & 31.11 & 26.00 & 0.059 & 30.75 & 30.19 & 0.006 \\
 numHelpProvidedAT & 15.27 & 14.15 & 0.009 & 14.70 & 12.87 & 0.018 \\
 numQuestionsAsked30D & 2.00 & 1.19 & 0.197 & 1.98 & 1.96 & 0.004 \\
 numHelpProvided30D & 0.81 & 0.46 & 0.068 & 0.78 & 0.63 & 0.029 \\
 numQuestionsAsked7D & 0.57 & 0.34 & 0.174 & 0.56 & 0.56 & $-$0.001 \\
 numHelpProvided7D & 0.22 & 0.12 & 0.063 & 0.21 & 0.17 & 0.028 \\
 tagAnswerRateAvg & 0.51 & 0.44 & 0.695 & 0.51 & 0.51 & 0.004 \\
\bottomrule\end{tabular}}%

\vspace{0.3em}
\parbox{0.9\textwidth}{\footnotesize\textit{Note.} SMD = standardized mean difference. The conventional threshold for acceptable balance is $|\text{SMD}| < 0.1$. All post-matching SMDs fall well below this threshold.}
\end{table}

\subsection{Analytical Strategy: Cox Proportional Hazards with Time-Varying Covariates}

We model the hazard of helping using a Cox proportional hazards model with time-varying covariates \citep{andersen_gill_1982}. The counting-process formulation of \citet{andersen_gill_1982} extends the original Cox model (\citep{cox_regression_1972} to settings with time-varying covariates and recurrent events, providing the large-sample theoretical foundation (including consistency and asymptotic normality of the partial likelihood estimator) that justifies our approach.

For each matched pair of observations (one treated, one control), we partition the 4-day observation window into intervals defined by the key event times: the start of the window, the posting of the question, and the arrival of the answer (the actual answer arrival for treated users or the synthetic transition for controls).

The hazard function is specified as:

\begin{multline}
\label{eq:cox}
h(t \mid \mathbf{X}(t)) = h_0(t) \exp\big(\beta_0 \cdot \text{treatment} + \beta_1 \cdot \text{phasePostQuestion}(t) + \beta_2 \cdot \text{treatedPostQuestion}(t) \\
+ \beta_3 \cdot \text{phasePostAnswer}(t) + \beta_4 \cdot \text{isTreatedActive}(t) \big)
\end{multline}

\noindent where $h_0(t)$ is the baseline hazard. \textit{treatment} is a time-constant indicator for treatment group membership (1 if the user's question received an answer, 0 otherwise); it absorbs stable between-group differences in baseline helping rates. The remaining covariates are time-varying indicators that switch on at the appropriate event times. The coefficient of primary interest is $\beta_4$, the hazard ratio associated with \textit{isTreatedActive}, which captures the additional increase in helping hazard for treated users after they receive an answer, beyond any change attributable to the act of asking a question itself.

To test the moderation by response time (Hypothesis 3), we extend the model with two additional terms:

\begin{multline}
\label{eq:cox_speed}
h(t \mid \mathbf{X}(t)) = h_0(t) \exp\Big(\boldsymbol{\beta}' \mathbf{X}(t) \\
+ \gamma \cdot \text{isTreatedActive}(t) \times \widetilde{\log}(\text{RT}) + \delta \cdot \text{treatedPostQuestion}(t) \times \widetilde{\log}(\text{RT}) \Big)
\end{multline}

\noindent where $\widetilde{\log}(\text{RT})$ denotes the standardized log response-time variable (see below), $\gamma$ captures the moderation of the post-answer treatment effect by answer speed, and $\delta$ captures the analogous moderation during the waiting period. Including the waiting-period interaction ensures that the $\gamma$ estimate for the post-answer phase is not confounded with the correlation between response time and waiting-period activity. A negative $\gamma$ would indicate that faster answers produce a larger reciprocity effect (consistent with H3), while a positive $\gamma$ would indicate the opposite.

\paragraph{Stratification by Tenure.} To test Hypothesis 2, we fit separate models for each of the seven tenure buckets. This allows the baseline hazard and all coefficients to vary freely across levels of user experience, providing a nonparametric view of how reciprocity changes over the user trajectory.

\paragraph{Identification.} Our identification strategy rests on the matched DiD logic embedded in the model. Propensity score matching \citep{rosenbaum_rubin_1983} ensures that treated and control observations come from users with comparable baseline characteristics, including tenure, activity levels, and question topic difficulty. This matching absorbs between-user differences that might otherwise confound the treatment effect. We do not include user fixed effects in the survival models. Because matching is performed across different users (not within-user), and because the Cox model already leverages within-observation temporal variation through the time-varying covariate structure \citep{andersen_gill_1982}, user fixed effects are not required for identification. The DiD structure, comparing within-observation changes before and after the answer across treatment groups, isolates the causal effect of receiving help from stable individual differences. The waiting period covariate (\textit{treatedPostQuestion}) absorbs the potential lift in helping behavior that treated users already show after merely asking a question but before receiving an answer. The treatment effect we estimate cannot be driven by this pre-treatment trend but must be driven by the answer itself.

\section{Results}
\label{sec:results}

\subsection{Descriptive Statistics}

The dataset spans 24,537,194 question-level observations from 3,872,841 unique users on Stack Overflow, with 50\% of questions receiving at least one answer. The matched analytic sample retains one treated and one matched control user per question, yielding a 50\% treatment rate by construction. Help-event counts are right-skewed: the mean is 0.18 events per observation window (SD\,=\,1.26), consistent with helping being a relatively rare behavior for any individual user in any given window. Among treated users, the median response time from question to first answer is 0.34 hours, though the distribution is heavily right-skewed (mean 5.09 hours, SD\,=\,17.09), reflecting a minority of questions that wait many hours or days for a response. The sample spans all levels of platform experience, with the largest strata being early-experienced users (1--6 months: $n=4{,}078{,}207$) and established users (1--3 years: $n=6{,}473{,}619$).

\begin{table}[H]
\caption{Descriptive Statistics}
\label{tab:desc_stats}
\centering
\begin{tabular}{@{}lrrrr@{}}
\toprule
\textbf{Variable} & \textbf{Mean} & \textbf{SD} & \textbf{Median} & \textbf{N} \\
\midrule
Questions (observations) & & & & 24,537,194 \\
Unique users & & & & 3,872,841 \\
\% questions with answer & 50.00\% & & & \\
\midrule
Help events per window & 0.18 & 1.26 & 0.00 & \\
User tenure (days) & 668 & 851 & 335 & \\
Response time (hours, treated) & 5.09 & 17.09 & 0.34 & \\
\midrule
\multicolumn{5}{@{}l}{\textit{Response time (hours, treated) by Tenure Bucket}} \\
\hspace{1em} $<$ 1 Week & 4.53 & 15.75 & 0.32 & 2,026,261 \\
\hspace{1em} 1 Week - 1 Month & 3.60 & 13.86 & 0.26 & 781,098 \\
\hspace{1em} 1 - 6 Months & 3.84 & 14.46 & 0.26 & 2,039,050 \\
\hspace{1em} 6 - 12 Months & 4.25 & 15.44 & 0.28 & 1,503,424 \\
\hspace{1em} 1 - 3 Years & 5.12 & 17.19 & 0.34 & 3,236,780 \\
\hspace{1em} 3 - 6 Years & 6.71 & 19.89 & 0.50 & 1,821,391 \\
\hspace{1em} $>$ 6 Years & 8.69 & 22.76 & 0.76 & 860,392 \\
\midrule
\multicolumn{5}{@{}l}{\textit{N (unique questions) by Tenure Bucket}} \\
\hspace{1em} $<$ 1 Week & & & & 4,052,645 \\
\hspace{1em} 1 Week - 1 Month & & & & 1,562,250 \\
\hspace{1em} 1 - 6 Months & & & & 4,078,207 \\
\hspace{1em} 6 - 12 Months & & & & 3,006,890 \\
\hspace{1em} 1 - 3 Years & & & & 6,473,619 \\
\hspace{1em} 3 - 6 Years & & & & 3,642,799 \\
\hspace{1em} $>$ 6 Years & & & & 1,720,784 \\
\bottomrule
\end{tabular}
\end{table}

\subsection{Descriptive Evidence: Help Rates and the Parallel Trends Assumption}

\begin{figure}[h]
\centering
\includegraphics[width=\textwidth]{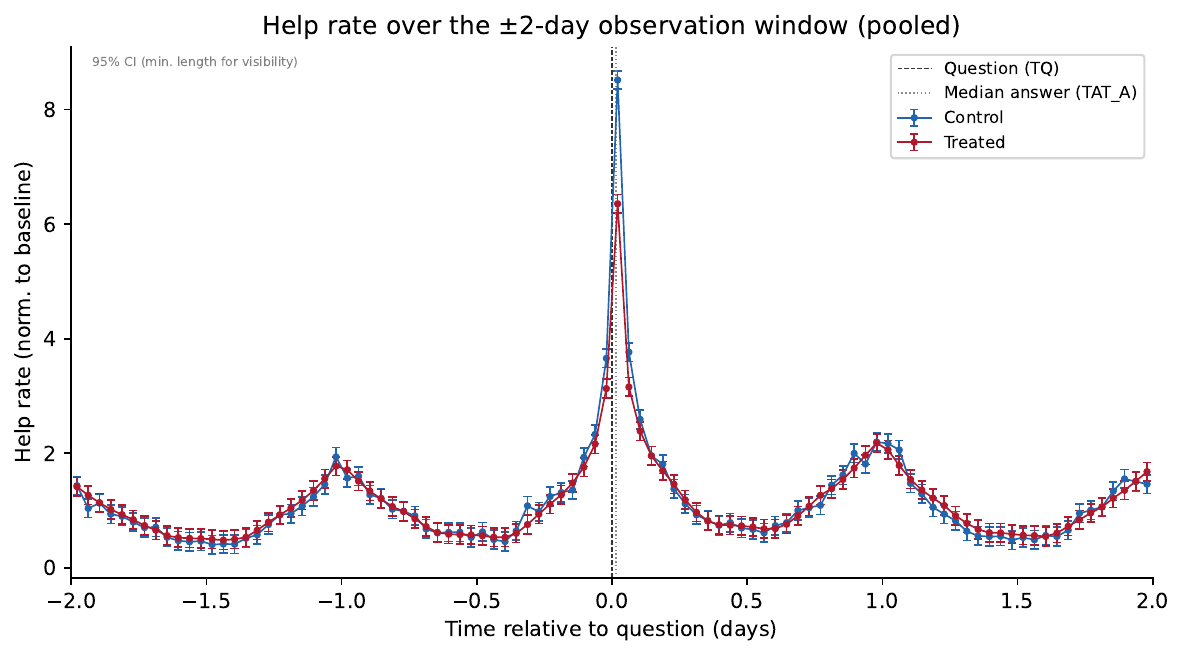}
\caption{Help rate over the $\pm$4-day observation window, pooled across all tenure buckets. The normalized help rate (relative to each group's pre-question baseline) is plotted for treated users (red, who received an answer) and matched control users (blue, who did not).}
\label{fig:help_rate_pooled}
\end{figure}

Figure~\ref{fig:help_rate_pooled} plots the normalized help rate for treated and control users over the two days before and after question posting. Before the question ($t < T_Q$), treated and control help rates track each other closely and exhibit no systematic divergence, consistent with the parallel trends assumption underlying the DiD design. After the median answer arrival time, treated users' help rate rises above that of controls, providing descriptive evidence of a reciprocity effect. The pre-question overlap also reflects the effectiveness of propensity-score matching: treated and control users enter the observation window with comparable baseline helping propensities. Notably, the control group spikes right after the question post time, which we will explain later. 

\subsection{Generalized Reciprocity Effect (H1)}

Table~\ref{tab:pooled_experienced} presents the pooled Cox model estimates across all experience levels. The coefficient on \textit{Received Answer $\times$ Post-Answer Received}, which captures the additional lift in helping hazard for treated users after receiving an answer relative to matched controls, is positive and highly statistically significant ($p < .001$). The estimated hazard ratio is 1.058 (95\%~CI\,=\,[1.04, 1.07]), indicating an approximately 5.8\% increase in the instantaneous rate of helping after receiving an answer. The waiting period coefficient (\textit{Received Answer $\times$ Post-Question}) is effectively zero ($-$0.005, $p > .10$), confirming that the lift in helping is attributable specifically to the arrival of the answer rather than to the act of asking itself. These results provide support for Hypothesis~1: receiving help increases a user's propensity to help others on the platform.

\begin{table}[H]
\caption{Pooled Cox Regressions (All Experience Levels)}
\label{tab:pooled_experienced}
\centering
\footnotesize
\begin{tabular}{@{}lcc@{}}
\toprule
 & \textbf{Main} & \textbf{Main + Response Time} \\
\midrule
\multicolumn{3}{@{}l}{\textit{Treatment Effect (DID)}} \\
\hspace{1em} Received Answer $\times$ Post-Answer Received & 0.0562*** & 0.0747*** \\
 & (0.0072) & (0.0062) \\
\hspace{1em} Hazard Ratio [95\% CI] & [1.04, 1.07] & — \\[4pt]
\multicolumn{3}{@{}l}{\textit{Waiting Period}} \\
\hspace{1em} Received Answer $\times$ Post-Question & -0.0050 & 0.0471*** \\[4pt]
\multicolumn{3}{@{}l}{\textit{Response Time Interaction}} \\
\hspace{1em} Treatment $\times$ log(Response Time) & — & -0.0447*** \\
 & — & (0.0021) \\[4pt]
\midrule
N & 24,537,194 & 24,537,194 \\
Events & 443,822 & 443,822 \\
\bottomrule
\multicolumn{3}{@{}l}{\footnotesize $^{***}p<0.001$; $^{**}p<0.01$; $^{*}p<0.05$; $^{\dagger}p<0.1$} \\
\end{tabular}
\end{table}

\subsection{Experience Moderation of Reciprocity (H2)}

\begin{figure}[h]
\centering
\includegraphics[width=\textwidth]{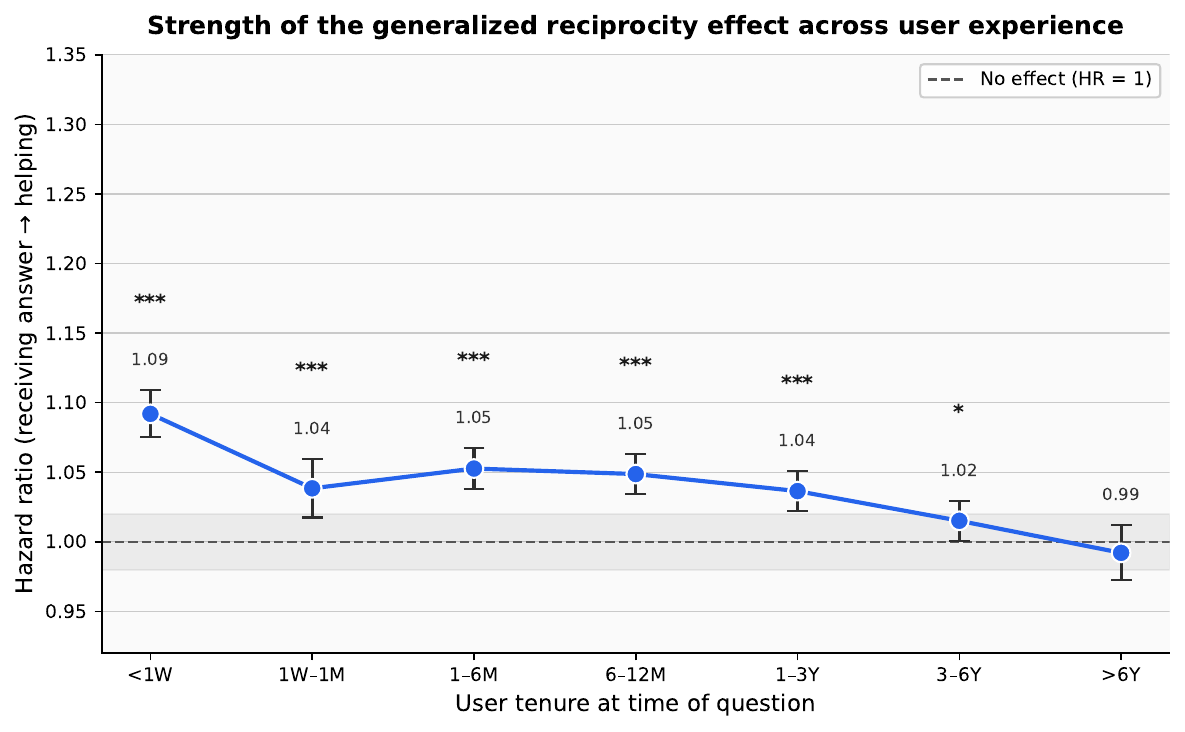}
\caption{Strength of the generalized reciprocity effect across user experience. Bars show the hazard ratio for the treatment effect by tenure bucket. Error bars indicate 95\% confidence intervals.}
\label{fig:strength_rec}
\end{figure}

To test Hypothesis~2, we estimate the Cox model separately for each of the seven tenure buckets. Table~\ref{tab:main_results} reports the estimates, and Figure~\ref{fig:strength_rec} plots the corresponding hazard ratios. The reciprocity effect is statistically significant in six of the seven tenure buckets, and its magnitude broadly declines with platform experience. The effect is largest among newcomers with less than one week of tenure (coefficient\,=\,0.0880, HR\,=\,1.09, 95\% CI\,=\,[1.08, 1.11], $p < .001$). The hazard ratio dips to 1.04 (95\% CI\,=\,[1.02, 1.06], $p < .001$) for users with one week to one month of tenure, before stabilizing through the intermediate experience strata (one to six months: HR\,=\,1.05, [1.04, 1.07]; six to twelve months: HR\,=\,1.05, [1.03, 1.06]; one to three years: HR\,=\,1.04, [1.02, 1.05]). The effect continues to decline for more established users (three to six years: HR\,=\,1.02, [1.00, 1.03], $p < .05$) and is statistically indistinguishable from zero among users with more than six years of tenure (coefficient\,=\,$-$0.0080, HR\,=\,0.99, [0.97, 1.01], $p > .10$). This gradient is visible in Figure~\ref{fig:strength_rec}. The waiting period coefficients remain near zero across all strata, ruling out pre-trend confounds. These results support Hypothesis~2: the reciprocity effect is strongest among newcomers, broadly attenuates with experience, and disappears entirely among the most experienced users.

\begin{landscape}
\begin{table}
\caption{Cox Regression Results: Effect of Receiving an Answer on Helping Hazard}
\label{tab:main_results}
\centering
\footnotesize
\begin{tabular}{@{}lccccccc@{}}
\toprule
 & $<$ 1 Week & 1 Week - 1 Month & 1 - 6 Months & 6 - 12 Months & 1 - 3 Years & 3 - 6 Years & $>$ 6 Years \\
\midrule
\multicolumn{8}{@{}l}{\textit{Treatment Effect (DID)}} \\
\hspace{1em}Received Answer $\times$ Post-Answer Received & 0.0880*** & 0.0377*** & 0.0513*** & 0.0475*** & 0.0358*** & 0.0150* & -0.0080 \\
 & (0.0079) & (0.0104) & (0.0071) & (0.0071) & (0.0070) & (0.0072) & (0.0102) \\
\hspace{1em}\textit{Hazard Ratio [95\% CI]} & [1.08, 1.11] & [1.02, 1.06] & [1.04, 1.07] & [1.03, 1.06] & [1.02, 1.05] & [1.00, 1.03] & [0.97, 1.01] \\[4pt]
\multicolumn{8}{@{}l}{\textit{Waiting Period}} \\
\hspace{1em}Received Answer $\times$ Post-Question & 0.0367*** & -0.0279** & 0.0063 & 0.0092 & 0.0104 & -0.0007 & -0.0096 \\[4pt]
\midrule
N & 4,052,645 & 1,562,250 & 4,078,207 & 3,006,890 & 6,473,619 & 3,642,799 & 1,720,784 \\
Events & 252,820 & 238,799 & 544,018 & 548,302 & 508,562 & 380,821 & 143,073 \\
\bottomrule
\multicolumn{8}{@{}l}{\footnotesize N = unique questions (treated + control) in the Cox sample. Within each column, treated vs.\ control counts can differ because tenure is defined per question.} \\
\multicolumn{8}{@{}l}{\footnotesize $^{***}p<0.001$; $^{**}p<0.01$; $^{*}p<0.05$; $^{\dagger}p<0.1$} \\
\end{tabular}
\end{table}
\end{landscape}

\subsection{Response Time Moderation (H3)}

Hypothesis~3 predicted that faster responses would strengthen the reciprocity effect. Table~\ref{tab:speed_results} reports the interaction between the treatment effect and log response time, estimated separately by tenure bucket. The results broadly support this prediction: the estimated speed interactions are negative across all tenure groups with a detectable treatment effect, indicating that faster responses are associated with stronger reciprocity.

Within the tenure-stratified Cox models, the speed interaction is negative across all groups with a significant base treatment effect, consistent with faster responses producing stronger reciprocity. The magnitude of this moderation varies with experience. Newcomers (tenure $<$\,1 week) show a moderate negative interaction ($\gamma = -0.0300$, SE\,=\,0.0024, $p < .001$). The interaction is largest in magnitude for users in the early-to-intermediate experience range: one week to one month ($\gamma = -0.0606$, SE\,=\,0.0029, $p < .001$) and one to six months ($\gamma = -0.0608$, SE\,=\,0.0020, $p < .001$). The interaction then attenuates with additional experience: six to twelve months ($\gamma = -0.0565$, SE\,=\,0.0020, $p < .001$), one to three years ($\gamma = -0.0383$, SE\,=\,0.0020, $p < .001$), and three to six years ($\gamma = -0.0189$, SE\,=\,0.0021, $p < .001$). Among users with more than six years of tenure, both the base treatment effect and the speed interaction are negligible and not statistically significant, consistent with the null result for this group in the main model.

\begin{figure}[h]
\centering
\includegraphics[width=\textwidth]{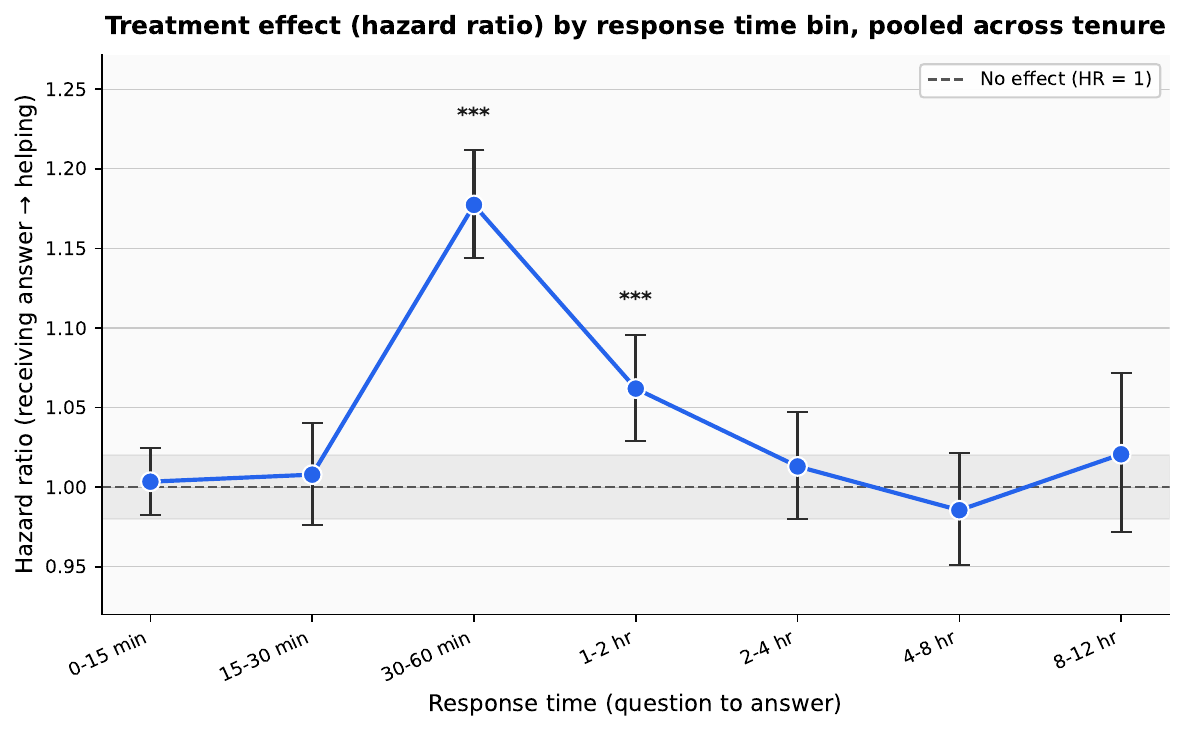}
\caption{Treatment effect (hazard ratio) by response time bin, pooled across all tenure buckets. Each point represents the estimated hazard ratio of the treatment indicator within that response-time bin, relative to the matched control. Error bars indicate 95\% confidence intervals.}
\label{fig:interaction_effect}
\end{figure}

Because we observed that the control group helped more than the treatment group immediately after the question time (Figure~\ref{fig:help_rate_pooled}), we suspect that users with very short response times will exhibit no treatment effect, potentially because some of them exit the platform immediately after receiving an answer. To probe this mechanism, Figure~\ref{fig:interaction_effect} plots the treatment effect (expressed as a hazard ratio) by discrete response time bins, pooled across all tenure groups. For the very fastest responses, the effect is indistinguishable from zero (0--15 min: HR\,$\approx$\,1.00; 15--30 min: HR\,$\approx$\,1.01; both $p > .10$). The effect peaks sharply for questions answered in 30--60 minutes (HR\,$\approx$\,1.18, $p < .001$) and declines thereafter: one to two hours (HR\,$\approx$\,1.06, $p < .001$), two to four hours (HR\,$\approx$\,1.02, $p > .10$), four to eight hours (HR\,$\approx$\,0.99, $p > .10$), and eight to twelve hours (HR\,$\approx$\,1.02, $p > .10$). This inverted-U, with its peak at 30--60 minutes rather than at the very fastest response bin, is visible in Figure~\ref{fig:interaction_effect}.

\begin{landscape}
\begin{table}
\caption{Response Time Moderation of the Reciprocity Effect}
\label{tab:speed_results}
\centering
\footnotesize
\begin{tabular}{@{}lccccccc@{}}
\toprule
 & $<$ 1 Week & 1 Week - 1 Month & 1 - 6 Months & 6 - 12 Months & 1 - 3 Years & 3 - 6 Years & $>$ 6 Years \\
\midrule
\multicolumn{8}{@{}l}{\textit{Base Treatment Effect}} \\
\hspace{1em}Received Answer $\times$ Post-Answer Received & 0.0606*** & 0.0675*** & 0.0820*** & 0.0800*** & 0.0675*** & 0.0348*** & 0.0020 \\
 & (0.0066) & (0.0088) & (0.0061) & (0.0061) & (0.0061) & (0.0063) & (0.0088) \\[4pt]
\multicolumn{8}{@{}l}{\textit{Response Time Interaction}} \\
\hspace{1em}Treatment $\times$ log(Response Time) & -0.0300*** & -0.0606*** & -0.0608*** & -0.0565*** & -0.0383*** & -0.0189*** & -0.0046 \\
 & (0.0024) & (0.0029) & (0.0020) & (0.0020) & (0.0020) & (0.0021) & (0.0031) \\[4pt]
\multicolumn{8}{@{}l}{\textit{Waiting Period $\times$ Response Time}} \\
\hspace{1em}Post-Question $\times$ log(Response Time) & 0.0125*** & 0.0164*** & 0.0261*** & 0.0258*** & 0.0224*** & 0.0160*** & 0.0101** \\[4pt]
\midrule
N & 4,052,645 & 1,562,250 & 4,078,207 & 3,006,890 & 6,473,619 & 3,642,799 & 1,720,784 \\
Events & 252,820 & 238,799 & 544,018 & 548,302 & 508,562 & 380,821 & 143,073 \\
\bottomrule
\multicolumn{8}{@{}l}{\footnotesize $^{***}p<0.001$; $^{**}p<0.01$; $^{*}p<0.05$; $^{\dagger}p<0.1$} \\
\end{tabular}
\end{table}
\end{landscape}

We interpret the non-linear pooled pattern through platform session dynamics. When a user posts a question, they enter a state of active, engaged waiting: they remain on the platform, browsing other questions, during the interval before their answer arrives. A very fast answer (under 30 minutes) resolves this pending state and can end the session --- some users simply leave after receiving help. Only users who have already left the platform can be re-engaged by the arrival of an answer. In the 30--60-minute range, many users have ended their session, but the question is still salient enough to draw them back when the answer arrives, creating a new session in which helping others becomes possible. At longer delays, the question's salience has faded, and the answer is less likely to prompt re-engagement at all. These dynamics produce the inverted-U in Figure~\ref{fig:interaction_effect}: the suppressed effect at very short response times reflects premature session closure rather than absent gratitude, and the attenuation at long delays reflects a question that no longer motivates return. The pattern is also consistent with the broadly negative continuous interactions, since many answers arrive after the 30-minute threshold, and with the relative spike in the control group just after question posting, since some treated users end their session early after receiving help. 

\begin{figure}
\centering
\includegraphics[width=\textwidth]{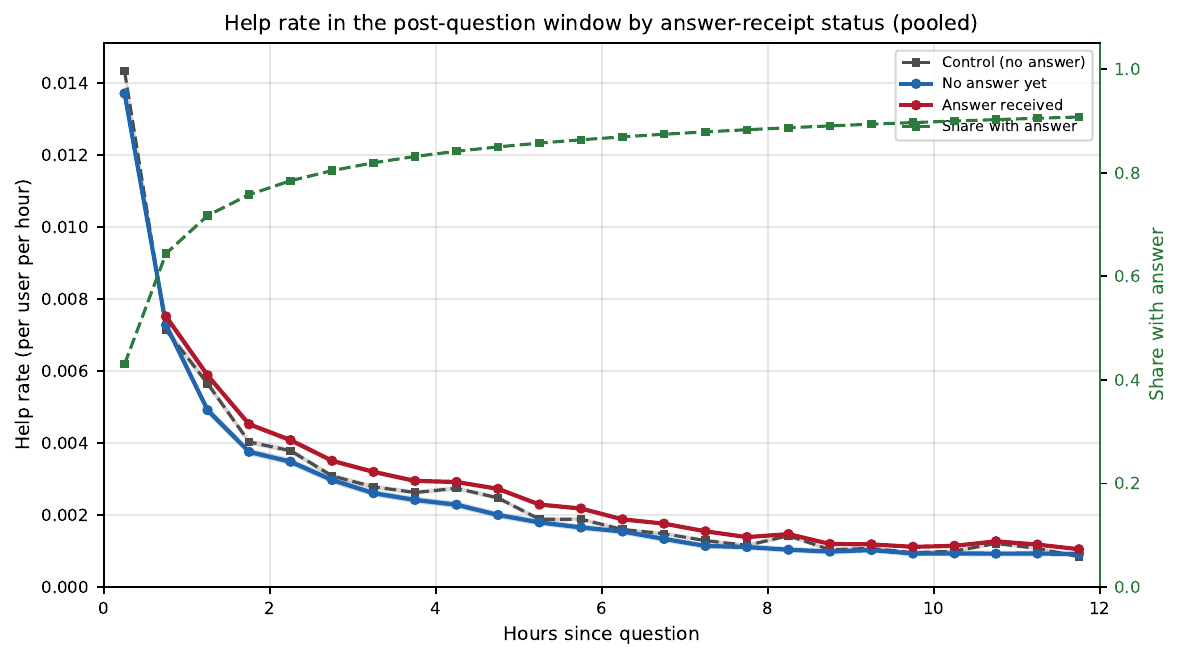}
\caption{Help rate in the post-question window by answer-receipt status, pooled across all tenure buckets. Grey dashed: matched control users who received no answer. Blue: treated users who have not yet received an answer (No answer yet). Red: treated users who have already received an answer (Answer received). Green dashed (right axis): cumulative share of treated users who have received an answer by each hour.}
\label{fig:help_rate_adoption_pooled}
\end{figure}

Figure~\ref{fig:help_rate_adoption_pooled} provides descriptive support. The plot compares users who have received an answer (red) and users who will receive one but are still waiting (blue). The gap between users who received help and those who still wait is small initially, then widens and converges at later responses, consistent with re-engagement being most likely in the intermediate response-time range. Tenure-stratified versions of this figure are in Appendix Figure~\ref{fig:help_rate_adoption_by_tenure}.

\section{Discussion}
\label{sec:discussion}

\subsection{Methodological Contribution}

A persistent difficulty in studying generalized reciprocity from platform data is that the same user characteristics that predict seeking help also predict giving it. Active, engaged users are more likely both to post questions and to post answers, so correlations between help received and help given can reflect differences in user engagement rather than any causal response to receiving help. Prior observational studies that report positive associations between help-seeking and subsequent helping behavior without accounting for this selection \citep{baker_paying_2014, chen_engaging_2018, joyce_predicting_2006} may have overstated the role of reciprocity. The matched DiD survival design developed here addresses this confound by comparing changes in helping behavior, before and after question posting, between users whose questions received answers and propensity-score-matched users whose questions did not. The waiting-period covariate further isolates the effect of answer receipt from the effect of question-posting itself, ruling out the possibility that the lift in helping reflects engagement created by the act of asking rather than by the act of receiving. Together, these features provide a more credible basis for causal inference than prior designs, and the estimated effect is substantially smaller than many earlier reports, consistent with the view that prior estimates were partially confounded.

\subsection{Reciprocity as a Contributor-Recruitment Mechanism}

The theoretical account developed in this paper treats generalized reciprocity not as a uniform driver of community cooperation, but as a mechanism whose salience varies with a user's position on the trajectory from newcomer to established contributor. The findings support this account. The reciprocity effect is present and meaningful among users early in their platform tenure, and it attenuates steadily across experience groups, becoming negligible among the most established users.

This gradient has implications for how reciprocity's role in online communities should be understood. If the effect were uniform across experience levels, reciprocity could plausibly be treated as a general engine of cooperative behavior, sustaining helping throughout a community's membership. The observed attenuation suggests a narrower role: reciprocity is most operative at the point of entry, providing an early push into helping before community-specific motivations are established. Newcomers lack the experience to value platform-specific incentives such as reputation accumulation or norm identification. The general moral impulse to reciprocate, what \citet{levine_logic_2020} terms the logic of universalization, is more prominent precisely because these community-specific motivations have not yet taken hold.

As users accumulate experience, this account predicts that community-specific incentives gradually displace the general reciprocal impulse. This is consistent with theories of norm displacement \citep{benabou_incentives_2006}, which show how the introduction of external incentives can crowd out moral motives. On Stack Overflow, reputation systems and status markers provide salient, community-specific reasons to help that are not available to newcomers. The absence of a detectable reciprocity effect among the most established users is consistent with this displacement account, though we cannot directly observe the motivational shift that the theory posits.

This reframing also helps account for the inconsistency of findings in the prior literature. Studies that draw primarily on samples of experienced users, or that aggregate across experience levels dominated by experienced contributors, should find weaker reciprocity effects under this account. Studies that capture behavior early in the user trajectory should find stronger ones. The variation in prior findings may thus reflect not noise or methodological failure, but genuine 
heterogeneity in the effect across the user population.

\subsection{Platform Session Dynamics and the Re-engagement Window}

The response time analysis qualifies the prediction, derived from laboratory-based theories of gratitude, that faster help should straightforwardly produce stronger reciprocity. The continuous interaction models are consistent with that prediction: across all experience groups with a detectable base effect, faster responses are associated with stronger reciprocity. The discrete-bin analysis, however, reveals a non-linearity in the pooled sample that the continuous model obscures. The reciprocity effect is negligible for answers arriving within the first thirty minutes, rises to its maximum in the thirty-to-sixty-minute range, and attenuates at longer delays. Emotional immediacy theory predicts a monotone decline from the fastest responses; the actual pattern is 
an inverted-U.

We interpret this as reflecting a re-engagement window created by the act of posting a question. When a user posts a question, they enter a state of active, engaged waiting: they remain on the platform, browsing other questions, during the interval before their answer arrives. An answer arriving very quickly, before this post-question engagement window has closed, resolves the pending state and can end the session --- some users simply leave after receiving help. Only users who have already ended their session can be re-engaged by the arrival of an answer. An answer in the thirty-to-sixty-minute range arrives after many users have left, but while the question is still salient enough to draw them back, creating a new session in which helping others becomes possible. At longer delays, the question's salience has faded, and the answer is less likely to prompt re-engagement at all. The theoretical implication is that gratitude alone is insufficient to predict reciprocal helping in field settings: what matters is whether the answer arrives in a window where it can re-engage a user who has left, rather than either closing an active session prematurely or arriving too late to be relevant.

This re-engagement mechanism has no natural analogue in laboratory settings, where participants remain present throughout an interaction. Laboratory designs that manipulate response speed isolate the emotional component of reciprocal motivation without the confound of session dynamics. Field settings introduce a structural layer: whether gratitude translates into helping depends not only on its intensity but on whether the answer arrives at a moment that brings the user back to the platform. Theories of generalized reciprocity developed in controlled settings need to incorporate this structural layer when applied to platform behavior.

The parallel between the attenuation in the base reciprocity effect and the attenuation in the 
speed interaction across experience groups is consistent with the norm-displacement account. As 
community-specific motivations take hold, they appear to weaken not only the general reciprocal 
impulse but also the sensitivity to response timing through which that impulse operates. The most 
experienced users show neither a base effect nor a speed moderation, suggesting that the mechanism 
as a whole has been displaced rather than merely dampened.

\subsection{Implications for Platform Design}

Because the reciprocity effect is concentrated among newcomers, the practical leverage that platforms can derive from it lies primarily in contributor recruitment rather than in sustaining contributions from established members.

The most direct implication is that platforms seeking to grow their contributor base should prioritize ensuring that new users receive answers to their questions. Answer receipt itself, rather than its speed, is the more important factor: the base effect is present across a wide range of response times, while speed moderation is secondary. Recommendation systems and visibility-boosting mechanisms that surface questions from new users to experienced answerers are, under this account, doing double work: they improve new users' direct experience and increase the probability that those users convert into contributors.

The re-engagement window finding points to a complementary consideration. Because re-engagement requires that the user has already left the platform, design features that artificially extend active sessions, such as surfacing related content during the waiting period, may actually reduce re-engagement opportunities by keeping users on the platform until after their answer arrives, at which point they leave without a reason to return. The more effective lever may be ensuring that answer notifications are salient enough to draw users back after they have left, preserving the re-engagement dynamic on which the peak reciprocity effect depends.

A further implication concerns the increasing prevalence of AI-generated answers on question-and-answer platforms. If AI assistance substitutes for human responses, the human-to-human interaction that creates the moral and emotional basis for generalized reciprocity may decline. Given that the effect is concentrated among newcomers, this substitution could impair the contributor-recruitment function that reciprocity appears to serve, reducing the pipeline of new contributors who currently receive an early push into helping from the experience of being helped by a human community member.

\subsection{Limitations}

Several limitations constrain the conclusions that can be drawn from this analysis.

The design identifies a behavioral response to answer receipt, but cannot identify the psychological mechanism producing it. Gratitude, moral obligation, and the re-engagement dynamics of platform sessions are all consistent with the observed patterns, and the data do not allow us to distinguish among them. Future work combining behavioral records with experience-sampling measures of emotional state or moral reasoning during platform use could help disentangle these pathways.

The identification strategy rests on the assumption that, conditional on the matched covariates and the DiD structure, answer receipt is approximately independent of the potential outcomes for helping behavior. The propensity score matching substantially improves the comparability of treated and control observations, as confirmed by post-matching balance diagnostics. Unobserved time-varying factors, including changes in workload, access to alternative information sources, or exogenous variation in a user's availability, could nonetheless bias the estimates in ways that the design cannot rule out.

Response time may partially proxy for question characteristics not fully captured by the tag-level answer rate, since questions that attract fast responses may differ systematically from those that wait. Future work with random treatment or finer-grained measures of question difficulty, such as text-based complexity or specificity scores, could provide a cleaner test of the response time mechanism.

The re-engagement window interpretation rests on indirect evidence: the non-linearity in the discrete-bin analysis, the broadly negative continuous speed interactions, and descriptive patterns in the help-rate figures. Direct tests would require session-level data, including login and logout timestamps or inactivity-based session boundaries, to confirm that the peak effect occurs among users who left and were drawn back by the answer notification, and that very fast answers disproportionately end active sessions. This remains an important direction for future work.

Stack Overflow is a large, relatively anonymous platform built around technical question-answering. The mechanisms identified here are likely to operate in other knowledge-sharing communities, but the specific patterns, including the tenure gradient, the shape of the re-engagement window, and the attenuation at high experience levels, may differ in smaller or more intimate settings, in intra-organizational platforms where social ties moderate motivation, or on platforms with different incentive structures.

Finally, the analysis establishes that reciprocity declines with experience but does not characterize what replaces it. The theoretical argument implicates com\-munity-spe\-cific norms and status motives, but testing this account would require integrating behavioral analysis with direct measures of those alternative mechanisms, for instance, through surveys or experimental manipulations of reputation visibility.

\section{Conclusion} \label{sec:conclusion} 
Generalized reciprocity is widely theorized as a mechanism sustaining cooperation on knowledge-sharing platforms, but observational evidence is limited by a persistent confound: active users are more likely both to seek help and to give it, regardless of whether their questions are resolved. Our matched DiD design addresses this confound. The effect is positive but more modest than prior reports suggest. Reciprocity is strongest among newcomers and disappears among the most established contributors, supporting a respecification of its role: not a general engine of community cooperation, but a recruitment mechanism that operates before platform-specific incentives displace the general moral impulse to reciprocate. The response time findings add a structural qualification: reciprocity peaks within a re-engagement window in which an answer arrives after a user has left the platform but while the question is still salient enough to draw them back. Answers arriving too quickly close an active session before re-engagement is possible; answers arriving too late find a user for whom the question no longer matters. Gratitude may be present in both cases, but whether it produces helping depends on the structural conditions of platform re-engagement --- a mechanism absent from laboratory settings that field theories of reciprocity need to incorporate. Platforms can leverage reciprocity at the point of entry, but not to sustain established contributors. They should prioritize ensuring newcomers receive answers and attend to the conditions that make re-engagement possible when those answers arrive.

\medskip
\bibliography{references.bib, references-2}

\newpage
\appendix

\section{Code for the paper}
Code for the paper can be found in the GitHub repo: \href{https://github.com/svenprue/prosocial-status-online-community}{svenprue/prosocial-status-online-community}

\section{Matching Diagnostics}
\label{sec:balance_plot}
\begin{figure}[!htbp]
    \centering
    \includegraphics[width=\textwidth]{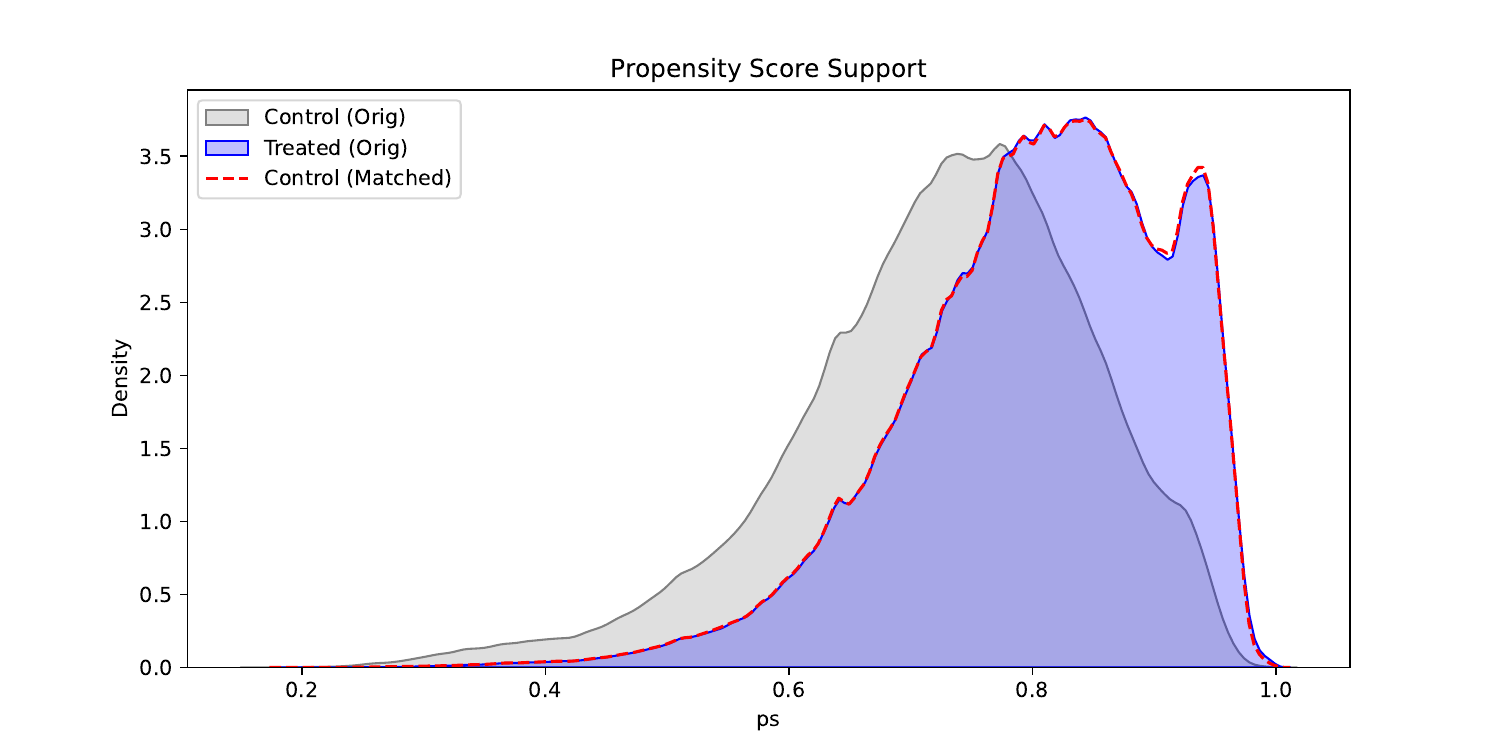}
    \caption{Distribution of propensity scores of treatment group, control group, and matched control group}
    \label{fig:common_support}
\end{figure}

Figure \ref{fig:common_support} shows that there is sufficient common support between the treatment and the control group. The propensity score distribution between the original treated and control groups is similar. The distribution of the matched control is nearly identical to the treated group. 

\begin{figure}[!htbp]
    \centering
    \includegraphics[width=\textwidth]{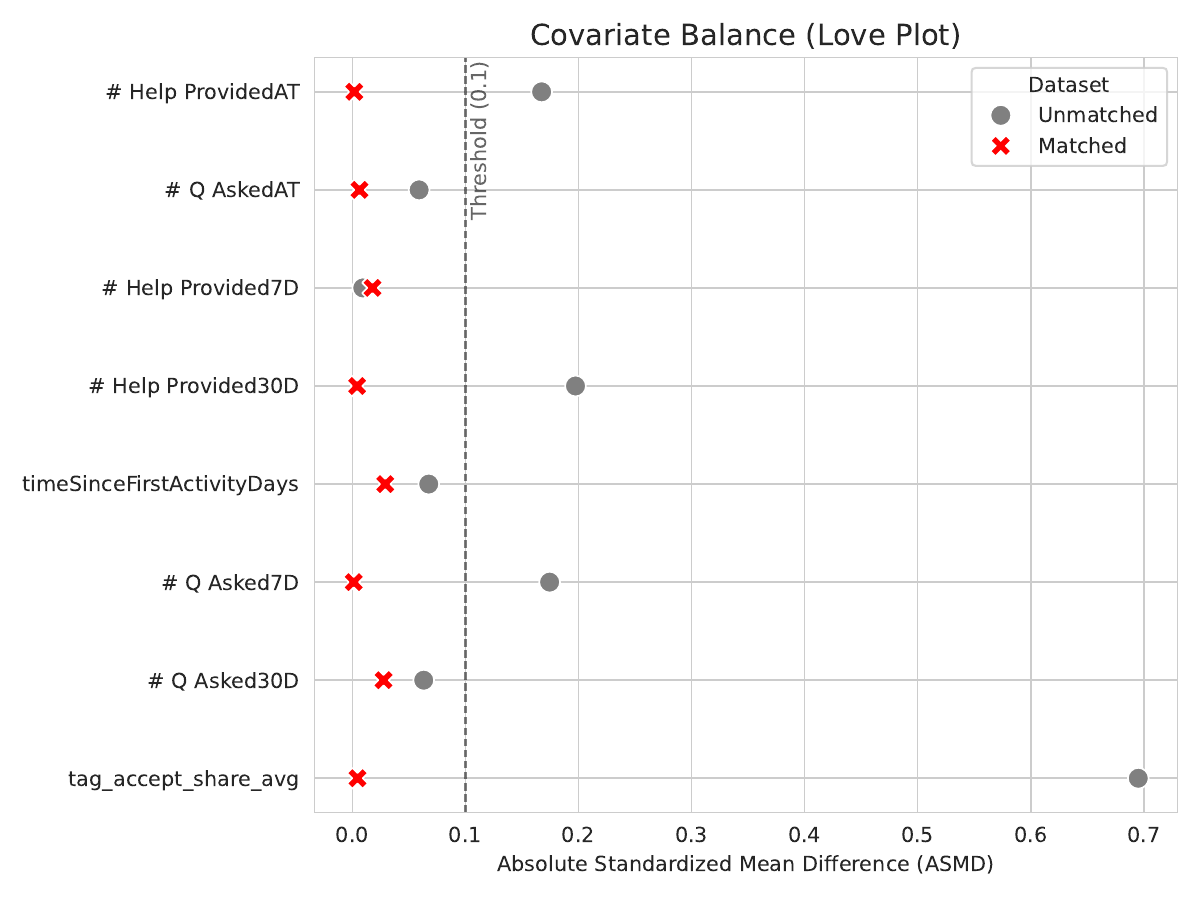}
    \caption{Visualization of balance in the unmatched and matched dataset}
    \label{fig:love_plot}
\end{figure}

Figure \ref{fig:love_plot} displays the standardized mean differences (SMD) between the matching variables before and after matching. While there are noticeable differences before the matching, all SMDs are very close to zero after the matching. 

\section{Help Rate Over the Observation Window by Tenure}
\label{sec:help_rate_by_tenure}

\begin{figure}[!htbp]
\centering
\includegraphics[width=\textwidth]{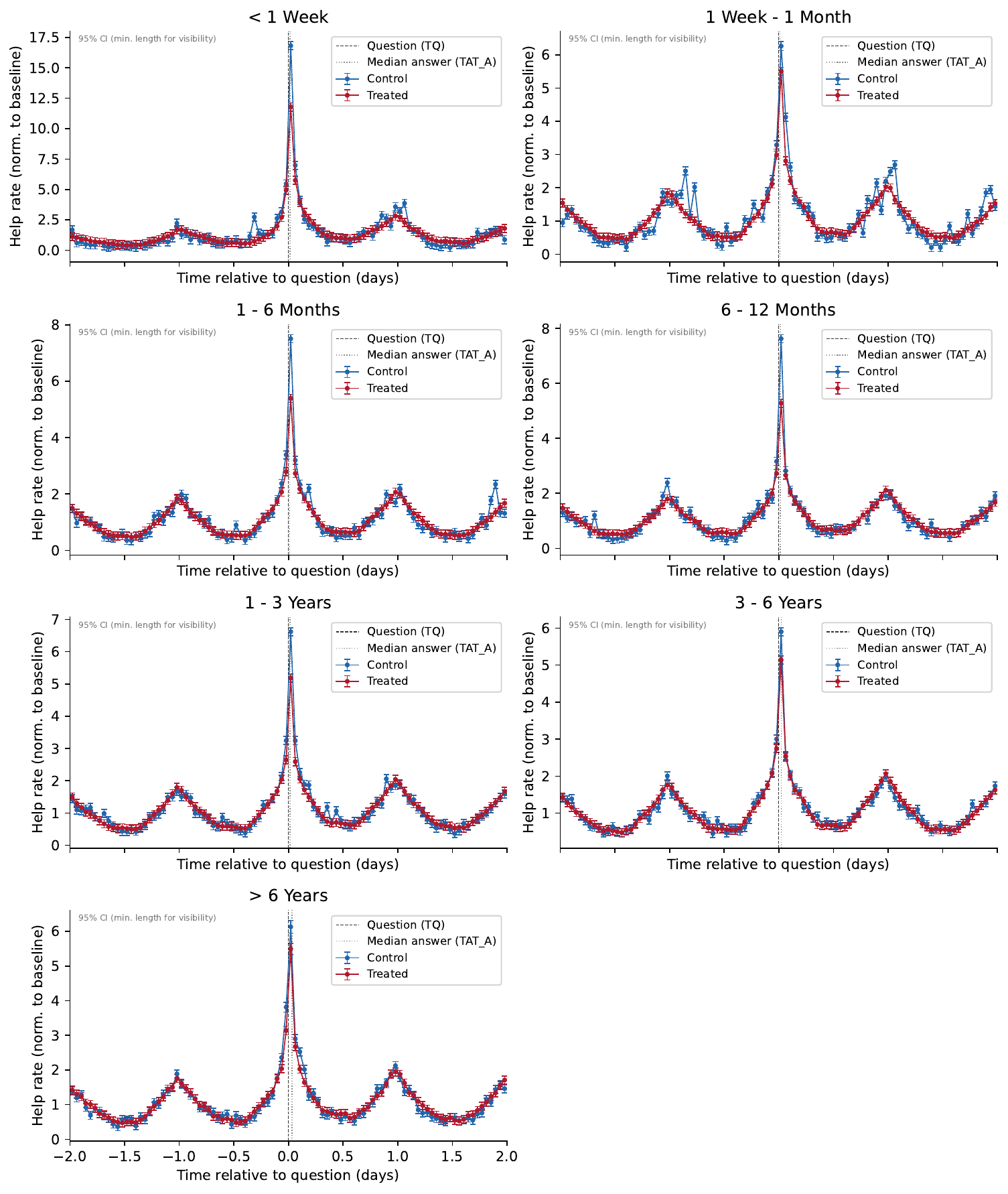}
\caption{Help Rate Over the Observation Window, by Tenure Bucket. Each panel plots the normalized help rate (relative to each group's pre-question baseline) for treated users (red) and matched control users (blue) over the $\pm$2-day window around the question-posting time ($T_Q$, dashed vertical line). The dotted vertical line marks the median answer arrival time ($T_A$). Error bars indicate 95\% confidence intervals.}
\label{fig:help_rate_by_tenure}
\end{figure}

Figure \ref{fig:help_rate_by_tenure} shows that the post-question lift in treated users' help rate is visible across all tenure buckets and is most pronounced among newcomers (tenure $<$ 1 week), consistent with the pattern reported in Table~\ref{tab:main_results} and Figure~\ref{fig:strength_rec}.

\section{Help Rate by Adoption Status and Tenure Bucket}
\label{sec:help_rate_adoption_by_tenure}

\begin{figure}[!htbp]
\centering
\includegraphics[width=.9\textwidth]{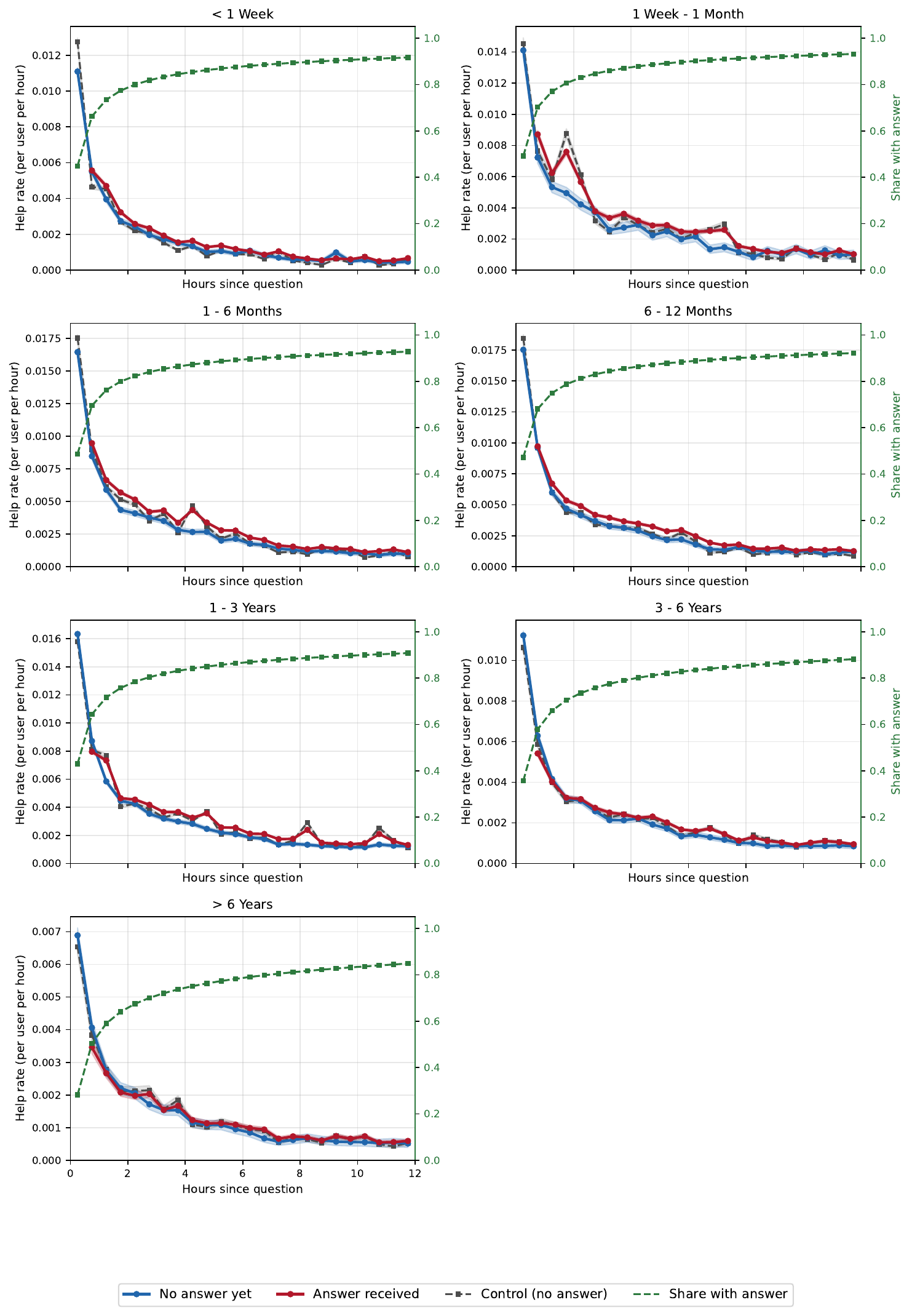}
\caption{Help rate in the post-question window by answer-receipt status, separately by tenure bucket. 
}
\label{fig:help_rate_adoption_by_tenure}
\end{figure}

Figure \ref{fig:help_rate_adoption_by_tenure} shows that for experienced users (all panels beyond $<$\,1 Week), the help rate of users still waiting (blue) exceeds that of users who have already received an answer (red), and both exceed the control baseline (grey dashed), up to approximately 1--2 hours post-question. This is consistent with the re-engagement window account: the pending-question state elevates engagement, and answer receipt collapses it back toward control levels. The newcomer panel ($<$\,1 Week) shows the reverse ordering, consistent with the gratitude-and-obligation mechanism dominating for users who do not promptly exit after receiving help.

section{Analysis of Life-Time Contributions \label{sec:lt-contributions}}
What remains unclear is whether users who become contributors through generalized reciprocity continue to contribute consistently and become valuable community members. To explore this, we compare the contributions of reciprocity-activated users with those of all contributors selected for Study 1. Specifically, we examine the number of times users from each group provided and received help up to July 14, 2024. The results are presented in Table \ref{tab:quartile_statistics}. Across all quartiles and at the median, reciprocity-activated users ask and provide more help than their counterparts.
For example, at the median, reciprocity-activated users provide eight answers and ask 13 questions, compared to six answers and 11 questions among all contributors. Similarly, the 75th percentile of reciprocity-activated users provides 23 answers and asks 29 questions, compared to 19 answers and 22 questions for all contributors.
These findings suggest that users who begin contributing after receiving help become equally, if not more, engaged in the community's helping dynamics compared to all contributors. Furthermore, this analysis demonstrates that once users start contributing, the act of giving and receiving help becomes (at least temporarily) decoupled.

\begin{table}[H]
\caption{Statistics on Contributions by User Sample}
\label{tab:quartile_statistics}
\centering
\begin{tabular}{lcccccc}
\toprule
 & \multicolumn{3}{c}{Answers Provided} & \multicolumn{3}{c}{Questions Asked} \\
\cmidrule(lr){2-4} \cmidrule(lr){5-7}
 & $Q_1$ & Med & $Q_3$ & $Q_1$ & Med & $Q_3$ \\
\midrule
Reciprocity-activated (\(N = 30,160\)) & 3.0 & 8.0 & 23.0 & 7.0 & 13.0 & 29.0 \\
All Contributors (\(N = 287,959\)) & 2.0 & 6.0 & 19.0 & 6.0 & 11.0 & 22.0 \\
\bottomrule
\end{tabular}
\end{table}

\section{Implementation Details} All models are estimated using the \texttt{lifelines} CoxTimeVaryingFitter with a small ridge (L2) penalizer applied for numerical stability. The penalizer strength is $\lambda = 5 \times 10^{-3}$ for main-effect models and $\lambda = 10^{-2}$ for models that include response-time interaction terms (where the interaction covariates are correlated with the base treatment indicator). These values are sufficiently small that they introduce negligible shrinkage relative to the scale of the estimated effects, but they ensure stable Hessian inversion in large risk sets. All covariates are mean-centered before fitting; response-time interaction terms are additionally clipped to their 5th--95th percentile range and z-scored as described above. Standard errors are computed from the penalized partial likelihood inverse Hessian without a sandwich (robust) correction. Because matched pairs are constructed across different users and the time-varying covariate structure eliminates the need for within-user corrections, this approach is appropriate; we note that clustering at the match level would be the natural extension if robust SEs were desired. For the pooled experienced-user models (H1), estimation was performed on a random sample of approximately 8,001,064 user-interval rows drawn from 3,306,638 matched pairs (preserving pair integrity) to keep computation feasible; tenure-bucket-specific models (H2, H3) use all available intervals within each stratum.

\end{document}